\documentclass[prb,a4paper,superscriptaddress,twocolumn]{revtex4}
\usepackage{amssymb}
\usepackage{amsmath}
\usepackage{amsfonts}
\usepackage{graphicx}
\usepackage{float}
\usepackage{bm}
\usepackage{color, comment}
\usepackage{hyperref}
\usepackage{comment}
\usepackage{physics}
\usepackage{ulem}

\usepackage{xcolor}

\begin{document}

\title{Quantum viscosity mechanism of the dissipative  dynamics in the Dicke model expressed via Lindblad equation of motion}
%\title{Energy dynamics in the Dicke model under influence of a 'bare' and 'dressed' Lindblad equation of motion}
\author{M.E.S. Beck}
\affiliation{National University of Science and Technology MISIS, Moscow, 119049 Russia}

\author{S.S. Seidov}
\affiliation{HSE University, Moscow 101000, Russia}

\author{S.I. Mukhin}
\affiliation{National University of Science and Technology MISIS, Moscow, 119049 Russia}

\date{\today}

\begin{abstract}
Quantum dissipation is studied in the superradiant phase of the Extended Dicke model. It is demonstrated analytically by quantum mechanical derivation of the Lindblad equation for the Dicke model in the superradiant state coupled to Caldeira-Leggett thermal bath, that the effective viscosity appearing in the semiclassical equations of motion of polaritonic condensate survives in the zero temperature limit $T\rightarrow 0$. The nonzero contribution to viscosity contains prefactor $n_B(\omega,T)+1$ with the Bose-Einstein function $n_B$ of harmonic oscillators in the thermal bath, indicating that virtual excitations of  harmonic oscillators in the thermal bath coupled to the polaritons of Dicke model give rise to effective viscosity in the $T\rightarrow 0$ limit. Besides,  it is demonstrated analytically, that correct expression for Lindbladian in the superradiant phase should be built using condensate-shifted creation and annihilation operators of the photons and pseudospin operators in Holstein-Primakoff representation of the coupled to photons two-level systems in order the system could relax to its minimum energy in $T\rightarrow 0$ limit due to thermal bath provided effective viscosity.
\end{abstract}

\maketitle

\section{Introduction}
From clouds of excited gas molecules to applications in quantum computing, the Dicke model has been applied to a wide range of fields where light-matter interactions play an important role. The model was initially used in order to study the spontaneous emission of radiation by a collection of excited molecules in free space ~\cite{dicke1954coherence, hepp1973superradiant}. The ability of the model to describe interactions between N atoms (modelled as two-level systems) and electromagnetic field made the model applicable to the fields of both Cavity ~\cite{de2018cavity, jaako2016ultrastrong, de2018breakdown, de2023relaxation} and Circuit Quantum Electrodynamics~\cite{mukhin2018first, mukherjee2020preparing, bhattacharjee2021quantum, dou2023superconducting, crescente2020ultrafast, seidov2024quantum}. In the former, prospects have been explored to apply the model to quantum assisted chemistry. Regarding circuit quantum electrodynamics, the Dicke model has been applied to, among others, quantum computing\cite{mukherjee2020preparing} and energy storage (Dicke Quantum Battery)\cite{bhattacharjee2021quantum, dou2023superconducting, crescente2020ultrafast, seidov2024quantum}. With the prospects of storing energy, the description of energy dissipation in the Dicke model has gained attention\cite{dou2023superconducting, kirton2017suppressing, boneberg2022quantum}.

The introduction of dissipation into quantum mechanical systems is frequently performed via the Gorini-Kossakowski-Sudarshan-Lindblad (GKSL) equation ~\cite{lindblad1976generators, gorini1976completely}. The GKSL equation is a modification of the Heisenberg equation of motion for an operator (or the Von Neumann equation of the time evolution for a density matrix) to include a dissipative non-unitary part, referred to in this article as the 'dissipator'. This dissipator describing the 'Heisenberg' equations of motion for an operator $\hat{O}$ has the form $\mathcal{D}_L[\hat{O}] = 2\hat{L}^\dagger \hat{O} \hat{L} - \{\hat{L}^\dagger \hat{L}, \hat{O}\}$, where $\hat{L}$ is the lowering operator and is typically chosen to be either the photonic ($\hat{a}$) or the lowering operator of the z-component of the superspin ($\hat{S}^-$). The introduction of dissipation into quantum systems in the ultra strong coupling (USC) regime encountered a problem related with the finding that quantum systems may spontaneously gain energy even at $T=0$. Studies of the Rabi model have shown that the Lindblad operator needs to be constructed from dressed operators (as opposed to the bare dissipator) as was shown in several pivotal works \cite{ridolfo2012photon, beaudoin2011dissipation, settineri2018dissipation}.  Quantum mechanically the explanation given why the system fails to relax to its ground state is that negative frequency modes enter into the dissipator \cite{frisk2019ultrastrong} preventing the system to go to its ground state. Despite this, the description of dissipation using a 'bare' dissipator still is a persistent phenomenon, including in the Dicke model\cite{dou2023superconducting, kirton2017suppressing, boneberg2022quantum}.

The Dicke model allows to study the effects of dissipation in the USC-regime semi-classically. Unlike the Rabi-model, which treats only one spin, the Dicke model treats a multitude of spins. This large number of spins allows to examine both the photonic field and the spins semi-classically. Previous works have examined dissipation in the Dicke model in the USC-regime. Amongst such studies are quantum path integral approaches\cite{liu2021finite} and application of numerical simulations using a Landau-Lifshitz-Gilbert approach confirming that indeed the Dicke model relaxes to its ground state even in the strong coupling regime\cite{del2024mumax3}. An importance of using 'dressed' Lindblad equations to describe the transmission spectura in a damped Dicke model was identified \cite{zueco2021photon}. 
 
In this article dissipation in the USC-regime will be studied semi-classcially in the Dicke Model in order to get a 'classical' insight into the breakdown of the Lindblad type equation constructed from 'bare' creation/annihilation operators.  In order to study dissipation, first, a 'bare' Lindblad equation will be constructed and, conforming with expectactions, it will be shown that such a Lindblad-operator can pump a (macroscopic) amount of energy into the system. Then, a dressed dissipator will be derived, corresponding to the spontaneously broken symmetry of the Dicke model in the USC-regime for large spins. It is shown that such a dressed dissipator does allow the system to relax to its ground state. Finally, a full derivation will be given of a quantum master equation of the Extended Dicke model coupled to a Caldeira-Leggett type bath of harmonic oscillators \cite{caldeira1981influence}. As a result, it is shown that effective viscosity entering the dissipator remains nonzero in the $T\rightarrow 0$ limit due to virtual excitations of  harmonic oscillators in the thermal bath coupled to the polaritons of the Dicke model. The associated dissipator allows to study the Extended Dicke Model in the (non-tunneling) spontaneously broken symmetry superradiant state for large total (pseudo)spins ($S=N/2\gg1$).

The article is structured as follows. In Section \ref{SectionHamiltonian} the Hamiltonian of the more general Extended Dicke Model is introduced and its phases are discussed. Section \ref{SectionDissipatorBare} presents the derivations of the equations of motion for dissipation described by a dissipator constructed from the bare spin and photonic operators. In addition, it will be shown that such a dissipator pumps energy into the system in the strong coupling regime. Section \ref{SectionShiftedHO} demonstrates, using the example of a shifted Harmonic Oscillator model, that if the position of the energy minimum is shifted but the dissipator remains unchanged, such a bare dissipator will no longer allow the system to evolve into the energy minimum. Finally, in section \ref{SectionDissipatorDressed}, a dissipator will be found by deriving a master equation. It will be shown that such a dissipator is built from the diagonal/dressed creation and annihilation operators and does allow the Dicke model to relax to its ground state. Corresponding expression for the viscosity of the Dicke model in the superradiant state will be derived.

\section{Model}

\subsection{The Hamiltonian}\label{SectionHamiltonian}
The Hamiltonian of the Extended Dicke model is given by:
\begin{gather}\label{EqHamiltonianEDM}
	\hat{H} = \frac{1}{2}\left(\hat{p}^2 + \omega^2\hat{q}^2\right) + g\hat{p}\hat{S}_y - E_Z \hat{S}_z + \left(1+\epsilon\right)\frac{g^2}{2}\hat{S}_y^2.
\end{gather}
Here $\hat{q}$ and $\hat{p}$ are respectively, the position and momentum operators of an electromagnetic field of frequency $\omega$. There is a Zeeman-term given by $-E_Z \hat{S}_z$, where $E_Z$ is the Zeeman energy. $\hat{S}_{x,y,z} = \frac{1}{2}\sum_{i=1}^N\hat{\sigma}_i^{x,y,z}$ represents the superspin of an ensemble of spins. The coupling between the bosonic field and the spin is regulated by the coupling constant $g$. And finally, there is a dipole-dipole term, regulated by the parameter $\epsilon$.

\subsubsection{Hamiltonian in the semi-classical approximation}\label{SectionClassicalHamiltonian}
The Extended Dicke model has two major phases: the normal phase and the superradiant state. A semi-classical consideration ($\hat{O} \to O$) of this Hamiltonian \cite{seidov2023bound,mukhin2021dicke,mukhin2023correspondence} gives insight into both phases. The lowest energy in the normal phase is obtained for the coordinates $q^N = p^N = S^N_x = S^N_y =0$ and $S^N_z = + N / 2$. It's associated energy is given by $E^N = - E_Z N /2$. As for the superradiant state: by setting $q = S_x =0$, and recognizing that the lowest energy is obtained for maximum spin one can express $S_y$ in $S_z$, and minimizing the total energy with respect to $p$ and $S_z$ one can find that the coordinates that minimize the energy in the superradiant phase are: $S_z^{SR} = -E_Z/\epsilon g^2$, $S_y^{SR} = \pm \sqrt{S^2-(S_z^{SR})^2}$ and $p^{SR} = \mp g\sqrt{S^2-(S_z^{SR})^2}$. The associated energy is:
\begin{gather}\label{EqEnergySRP}
	E^{SR} = -E_Z S_z^{SR} + \frac{\epsilon g^2}{2}\left(S^2-(S_z^{SR})^2\right).
\end{gather}

Comparing the energies of the normal and superradiant phases, one is able to deduce that the superradiant phase becomes energetically favorable when $\epsilon < 0$ and $g > g_c = \sqrt{-E_Z/\epsilon S}$. It is remarked that the line connecting center of the sphere with the position on the spin sphere associated with minimum energy, makes  an angle with the $S_z$-axis, given by:
\begin{gather}\label{EqAngelTheta}
	\cos(\theta) = -E_Z/\epsilon g^2 S.
\end{gather}

\begin{figure}[t!]
	\centering
	
	\includegraphics[width=0.4\textwidth]{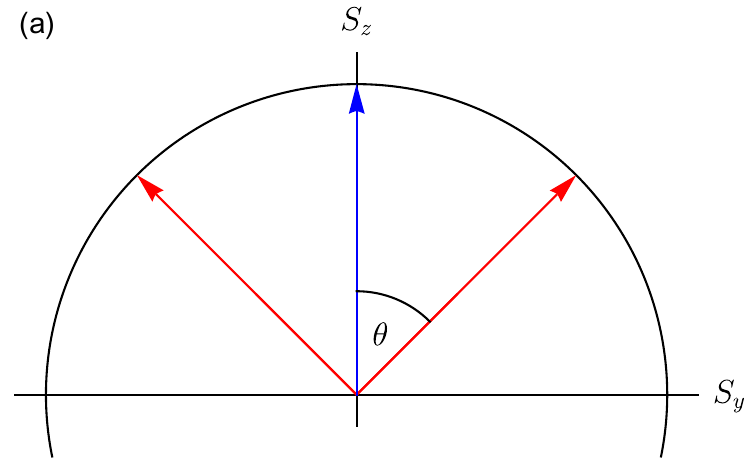}
	
	\includegraphics[width=0.4\textwidth]{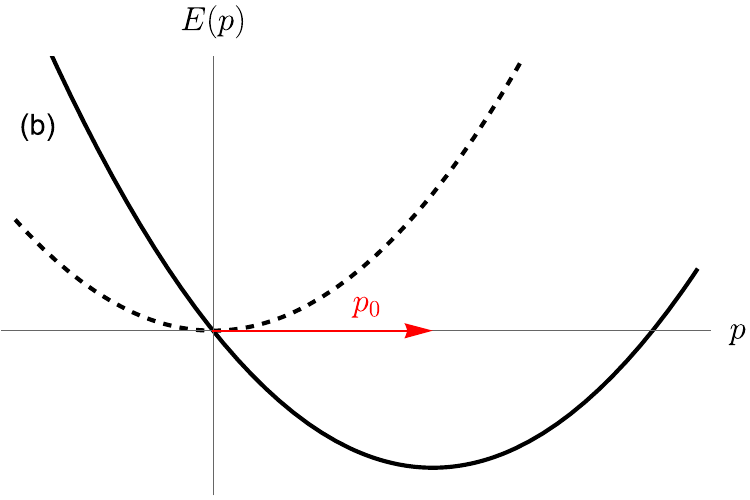}	
	
	\caption{Schematic depiction of the rotated and shifted minimum energy coordinates of the Extended Dicke model on the superspin sphere and the p-coordinate. (Fig.\ref{FIG_classicalEDM_shift_rotation}a) The blue arrow indicates the superspin coordinate of the energy minimum in the normal phase / weak coupling regime on the superspin sphere. Red arrows depict the coordinates of the (degenerate) minima energies in the superradiant / strong coupling regime. Compared to the minimum energy coordinate of the normal phase the minimum energy coordinates of the superradiant phase are rotated by an angle $\theta$.
(Fig.\ref{FIG_classicalEDM_shift_rotation}b) Energies associated to the (semi-classical) momentum-dependent part in the Hamiltonian ($p^2/2 - g p S_y$) at two different, but fixed, $S_y$. The dotted curve represents the momentum related energy when $S_y=0$ (the coordinate of minimum energy for $S_y$ in the normal phase) . The blue line represents momentum related part of the energy when $S_y \neq 0$ (as would be the case when the system would be in the superradiant phase). At minimum energy in the superradiant phase the momentum would be shifted by an amount $p_0$.}\label{FIG_classicalEDM_shift_rotation} 
\end{figure}

\subsection{Dissipator built from the bare operators: $a$, $S^+$}\label{SectionDissipatorBare}
This section derives the equations of motion for the operators $\hat{q}, \hat{p}, \hat{S}_{x,y,z}$ in the Heisenberg picture in the presence of dissipation using a bare Lindblad equation. These equations of motion will be studied semi-classically following similar derivations as done in \cite{seidov2023bound}. Here, the dissipator will be built from the bare bosonic- and superspin lowering operators ($\hat{a}$, $\hat{S}^-$).

In the Heisenberg picture the time evolution of an operator $\hat{O}$ under influence of dissipation, is given by the Lindblad equation:
\begin{gather}\label{EqLindbladDefinition}
	\dot{\hat{O}} = \mathcal{L}_L[\hat{O}] = i\left[\hat{H},\hat{O}\right] + \kappa \mathcal{D}_L[\hat{O}],
\end{gather}

where $\kappa$ is the viscosity constant and $[\hat{H},\hat{O}]$ is the commutator. The quantity $\mathcal{D}_L[\hat{O}] = 2\hat{L}^\dagger \hat{O} \hat{L} - \{\hat{L}^\dagger \hat{L}, \hat{O}\}$, will be referred to as the 'Dissipator' as it describes the non-unitary dynamics of the system. The dissipator in the Dicke model is constructed using the lowering (and raising) operators, in Eq.(\ref{EqLindbladDefinition}) the operator $\hat{L}$. When describing the Dicke model in the presence of dissipation, the dissipator is sometimes constructed \cite{boneberg2022quantum,dou2023superconducting, kirton2017suppressing} from the different bare raising and lowering operators; the bosonic ($\hat{a}$) and raising/lowering operators ($\hat{S}^\pm$) of z-component of the superspin. (Please note, that because of the minus sign in front of the Zeeman-term in the Hamiltonian, it is the $\hat{S}^+$-operator that lowers the energy.) Because there are two distinct dissipation channels (one bosonic, one spin), two different dissipators are obtained corresponding to $\kappa_1 \mathcal{D}_a[\hat{O}]$ (bosonic) and $\kappa_2 \mathcal{D}_{\hat{S}^+}[\hat{O}]$ (superspin) respectively. Using the Lindblad equation, formed by these 'bare' dissipators one can find the Heisenberg equation of motion for the prior mentioned operators to be:
\begin{gather}\label{EqEOMBare}
    \begin{split}
        \dot{\hat{q}} & = \hat{p} + g\hat{S}_y - \kappa_1 \hat{q} \\
	    \dot{\hat{p}} & = -\omega^2 \hat{q} - \kappa_1 \hat{p} \\
	    \dot{\hat{S}}_x & = g\hat{p}\hat{S}_z + E_Z \hat{S}_y - \kappa_2 \hat{S}_x + \left(1+\epsilon \right)\frac{g^2}{2}\{\hat{S}_y, \hat{S}_z\}\\ 
	    \dot{\hat{S}}_y & = -E_Z \hat{S}_x - \kappa_2 \hat{S}_y \\
	    \dot{\hat{S}}_z & = -g\hat{p}\hat{S}_x + \kappa_2 \left(N - 2\hat{S}_z\right) - \left(1+\epsilon\right)\frac{g^2}{2}\{\hat{S}_x,\hat{S}_y\}.
    \end{split}
\end{gather}

These equations of motion will be studied under the semi-classical approximations. It was pointed out {\cite{mukhin2023correspondence,mukhin2021dicke}} that in the large superspin limit ($S \gg 1$) the motion of the Dicke model in the superradiant phase can be described semi-classically. Hence, we introduce the semi-classical approximation by replacing quantum mechanical operators by c-numbers ($\hat{O} \to O$). Under the semi-classical approximation, the anti-commutator simplifies to $\{\hat{S}_\alpha,\hat{S}_\beta\} \to 2S_\alpha S_\beta$.

The fixed points of the resulting system of differential equations can be found by setting the time derives equal to zero. By doing so, three fixed points are found. The first, the trivial fixed point, is given by: $q_{stat,\pm}^{N} = p_{stat,\pm}^{N} = S_{x,stat,\pm}^{N} = S_{y,stat,\pm}^{N} = 0$ and $S_{z,stat}^N = N/2$, corresponding to the minimum in energy of the normal phase. The other two points are fixed points only in the superradiant phase. They are given by:
\begin{gather}\label{EqFixedPointsBare}
    \begin{split}
            q_{stat,\pm}^{SR} & = - \frac{\kappa_1}{\omega^2}p_{stat,\pm}\\
            p_{stat,\pm}^{SR} & = -\frac{g}{u}S_{y,stat,\pm}\\
            S_{x,stat,\pm}^{SR} & =  -\frac{\kappa_2}{E_Z}S_{y,stat,\pm}\\
            S_{y,stat,\pm}^{SR} & =\pm\sqrt{S_{z,stat}\left(N-2S_{z,stat}\right)/v}\\
            S_z^{SR} & = \frac{E_Z}{g^2}\frac{uv}{1-(1+\epsilon)u},
    \end{split}
\end{gather}

here $u = 1 + \kappa_1^2/\omega^2$ and $v = 1 + \kappa_2^2/E_Z^2$. Additionally attention is drawn to the following: the latter two fixed points are only fixed points when $g$ and $\epsilon$ satisfy the conditions $\epsilon \leq \epsilon_c$ and $g \geq g_c$. These critical values are given by $g_c = \sqrt{\frac{2E_Z}{N}\frac{uv}{1-(1+\epsilon)u}}$ and $\epsilon_c = \frac{1}{u}-1$.

A disturbing phenomenon occurs when these fixed points are plugged into the Hamiltonian in order to find the corresponding energies. No matter what the value of any of the constants (coupling and viscosity constants included), all the fixed points always correspond to the same energy: $H_{EDM}[\mathbf{q}_{stat}^{N,SR}] = - E_Z N/2$. This energy corresponds with the ground state energy of the normal phase in model without dissipation. In the superradiant phase the system is able to lower its energy further by creating a dipole moment ($S_y > 0$, $p \neq 0$). However, the above results show that if a dissipator is added to the equations of motion constructed from the bare $\hat{a}$ and $S^-$ the system does not relax to its lowest energy for $\epsilon \leq \epsilon_c$, $g > g_c$.

\begin{figure}
	\centering
	
	\includegraphics[width=0.4\textwidth]{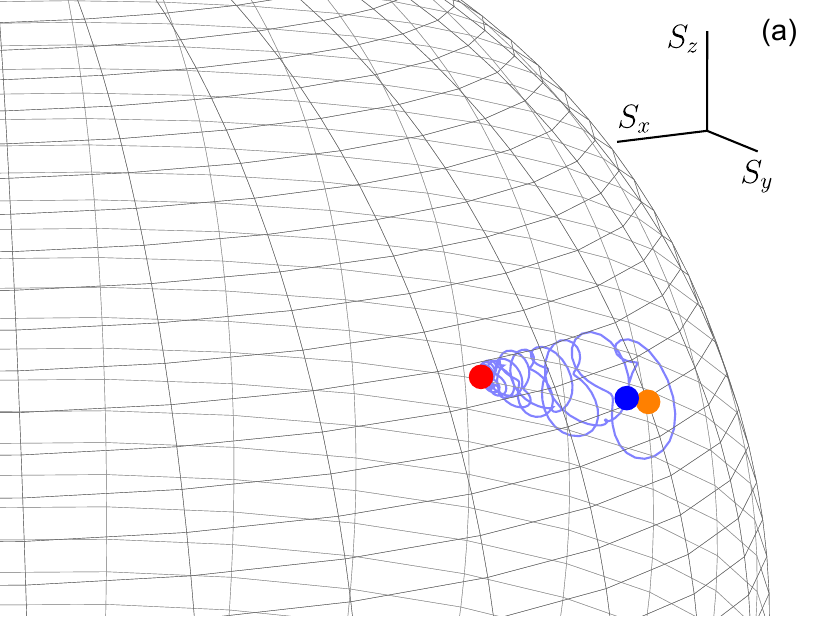}

	\includegraphics[width=0.4\textwidth]{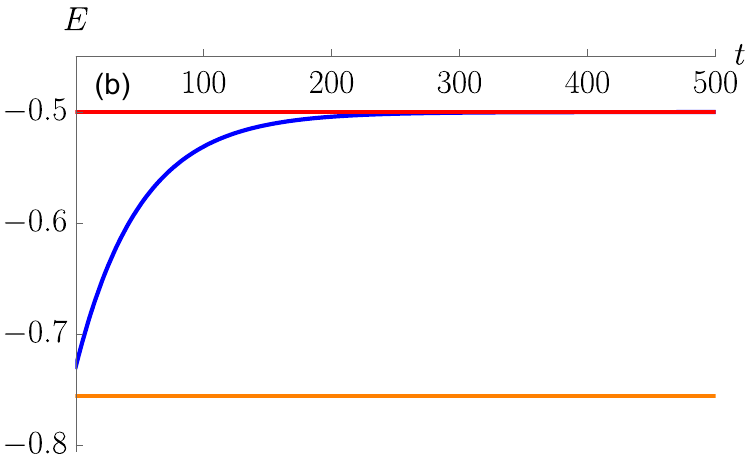}
	
	\caption{The semi-classical dynamics of the energy and spin coordinates of the Extended Dicke model under influence of a bare dissipator. (Parameters: $\epsilon=-1, \omega=1,E_Z=0.2,g=0.46, \kappa_1=\kappa_2=0.02$). Here (Fig.\ref{FIG_asigma_motion}a) demonstrates the dynamics of the $S_{x,y,z}$-coordinates on a superspin sphere. The orange points represent the coordinates of the minimum energy of the semi-classical approximation of the Extended Dicke model. The red points represent the fixed points of the Lindbladian dynamics described by the Heisenberg equations of motion with $\mathcal{D}_a[O]$ and $\mathcal{D}_{S^+}[O]$-dissipators. The blue point represents the starting coordinates of the dynamics described by the system of equations Eq.(\ref{EqEOMBare}). The initial coordinates are chosen such that they are close to the minimum in energy. The second graph (Fig.\ref{FIG_asigma_motion}b) represents the dynamics of the energy corresponding to the same trajectory, the blue curve. The orange line represents the minimum energy of the semi-classical Dicke model. The red line represents the energy corresponding to fixed points of the Lindbladian dynamics described by the bare $\mathcal{D}_a[O]$ and $\mathcal{D}_{S^+}[O]$-dissipators. One can see that the condensate gains energy due to the influence of the 'bare' dissipators}\label{FIG_asigma_motion} 
\end{figure}

To demonstrate that the Extended Dicke model does not relax to its ground state a numerical solution of the system of equations Eq.(\ref{EqEOMBare}) is presented in Fig.\ref{FIG_asigma_motion}. The starting position of the system (blue dot) is chosen to be close to one of the minima in energy of the semi-classical approximation of the Extended Dicke model (orange dots). The system moves away (blue line) from the minimum in energy towards one of the two fixed points (red dots, given by Eq.(\ref{EqFixedPointsBare})). The associated dynamics of the energy is plotted. The system moves away from the orange line (minimum energy), and grows ever closer to the red line (energy associated to the fixed point). As can be seen, the system actually gains energy (instead of losing it) when it starts close to the energy minimum due to the presence of the bare dissipator. This energy can be up to the order of $\propto N/2$.\\

\subsubsection{Dissipator of a shifted Harmonic Oscillator}\label{SectionShiftedHO}
In the prior section it was shown that using a 'bare' dissipator in the Extended Dicke model leads to the peculiar result that the system no longer relaxes to the coordinates of minimum energy in the strong coupling regime. This section will argue that the reason why the Extended Dicke model no longer relaxes to its ground state is because the bare dissipator fails to take into account a shift in $p$-coordinate and a rotation on the superspin sphere (as was mentioned in Section \ref{SectionClassicalHamiltonian}) of the minimum-energy position. This will be demonstrated by introducing a similar shift by $p_0$ of $p$-coordinate in the model of a simple harmonic oscillator. The dissipator will be constructed using: i. the  unshifted creation operator ($\mathcal{D}_a[\hat{O}]$) and ii. the shifted creation operator ($\mathcal{D}_{a-ip_0/\sqrt{2\omega}}[\hat{O}]$). It will be demonstrated that the former leads to a fixed point of the equations of motion that does not correspond to the new minimum energy. The fixed point of the 'shifted' dissipator does correspond to the minimum in energy.

As was shown in section \ref{SectionClassicalHamiltonian}, when the system enters the strong coupling regime, e.g. the superradiant phase, the equilibrium position of the moment $p$ and the superspin coordinates $J_y$ and $J_z$ change. In other words, the position of the minimum energy in phase space has shifted. Such a shift in phase space can be studied by a Harmonic Oscillator model. For simplicity, consider a Harmonic Oscillator with its (semi-classical) minimum located at $q=0,p=0$. The Hamiltonian, with creation/annihilation operators as defined in Section \ref{SectionHamiltonian}.
\begin{gather}
	\hat{H} = \frac{1}{2}\left(\hat{p}^2 + \omega^2\hat{q}^2\right) = \omega \hat{a}^\dagger \hat{a} + \frac{\omega}{2}.
\end{gather}

A shift is introduced in the momentum operator ($\hat{p} \to \hat{p} - p_0$, $\hat{q} \to \hat{q}$), similar to what happens with the Extended Dicke model when it enters the strong coupling regime. Consequently, the Hamiltonian changes to $\hat{H} \to \frac{1}{2}(\hat{p}-p_0)^2 + \frac{1}{2}\omega^2\hat{q}^2$ and has its (semi-classical) minimum at $q=0, p=p_0$. The associated shifted annihilation operators are: $\hat{a} \to \hat{a} - i p_0 / \sqrt{2\omega}$. Using the Heisenberg equations of motion and building the dissipator using the unshifted annihilation operators ($\mathcal{D}_a[\hat{a}]  = -\kappa \hat{a}$ and $\mathcal{D}_a[\hat{a}^\dagger]  = -\kappa \hat{a}^\dagger$), one obtains the following equations of motion:
\begin{gather}
    \begin{split}
        \dot{\hat{p}} = -\omega^2 \hat{q} - \kappa \hat{p}\\
	    \dot{\hat{q}} = \hat{p} - p_0 - \kappa \hat{q}.
    \end{split}
\end{gather}

If one were to examine these equations of motion semi-classically ($\hat{q} \to q$, $\hat{p} \to p$) one finds that the unitary parts of the equations of motion (found by setting $\kappa = 0$) still produce an expected result. However, when switching on dissipation ($\kappa > 0$) and eliminating $p$ gives the following dissipative equation of motion for the coordinate $q$:
\begin{gather}
	\ddot{q} + 2\kappa\dot{q} + (\omega^2 + \kappa^2)q + \kappa p_0 = 0
\end{gather}

Setting $\ddot{q}=0$, $\dot{q}=0$, one finds the fixed point of this equation to be:
\begin{gather}
    \begin{split}
        q_{fx} & = - \frac{\kappa}{\omega^2+\kappa^2}p_0\\
	    p_{fx} & = \frac{p_0}{1+(\kappa/\omega)^2},
    \end{split}
\end{gather}
I.e. the fixed point does not correspond to the minimum in energy of the semi-classical approximation of the Hamiltonian ($q=0$, $p=p_0$).

However, this problem is not encountered if the creation/annihilation operators, used to built the disspators, are also shifted. Using the shifted operators ($\tilde{a} = \hat{a} - ip_0/\sqrt{2\omega}$) to build the dissipator, leads to $\mathcal{D}_{\tilde{a}}[\hat{\tilde{a}}] = -\kappa \hat{\tilde{a}}$, $\mathcal{D}_{\tilde{a}}[\hat{\tilde{a}}^\dagger] = -\kappa \hat{\tilde{a}}^\dagger$ (As can easily be checked using the commutation relations $[\hat{\tilde{a}},\hat{\tilde{a}}^\dagger] = 1$). One obtains the quantum equations of motion in the form:
\begin{gather}
    \begin{split}
	   \dot{\hat{p}} &= -\omega^2 \hat{q} - \kappa (\hat{p}-p_0)\\
	   \dot{\hat{q}} & = \hat{p} - p_0 - \kappa \hat{q}.
    \end{split}
\end{gather}

When examined semi-classically, these equations of motion do let the system relax to the expected, shifted, position for the minimum energy ($q=0$, $p=p_0$) as required. Indeed, one sees in this case that failing to account for a shift when constructing the dissipator leads to a deviation in $q$ and $p$ that is proportional to $p_0$.

The connection to the Extended Dicke model is this; as with the shifted harmonic oscillator, the minimum energy $p$-coordinate of the Extended Dicke model is also shifted in the strong coupling regime (superradiant phase). Hence, not taking into account the shift in $p$-coordinate of the energy minimum when constructing the dissipator leads to a system that does not relax to its minimum in energy. In fact, to let the system relax to its energy minimum, all dissipators need to be associatively shifted in $p$-space and rotated on the spinsphere by an angle $\theta$. 

The phenomenon that a system does not relax to its ground state under influence of a bare dissipator is a general phenomenon and not just limited to the Extended Dicke model. If one of the positions or momenta associated to the minimum energy shifts when going from the weakly coupled phase to the strongly coupled phase, the system should not be expected to relax to its ground state energy under influence of a bare dissipator. Rather, a dissipator needs to be constructed of shifted and/or rotated creation annihilation operators at the very least to describe the system relaxing to the proper energy (though the dynamics in phase space is unlikely to be correct). Or more correctly, a dissipator needs to be derived via, for example, a Born-Markovian description of finding a master equation. It will be shown in the section \ref{SectionDissipatorDressed}) and the appendix that such a treatment leads to dissipators built from the diagonalized or dressed creation/annihilation operators.

Although it will not be demonstrated here, incorporating such a shift and translation 'ad hoc', does allow the system to relax to its minimum energy. Rather than an 'ad hoc' repair of dissipators, this article will present a formal derivation of the dissipator in the next section. The formal derivation of the dissipator is performed by deriving a quantum master equation in the strong coupling regime starting from weakly coupling the system to a Caldeira-Leggett bath.

\subsubsection{Ad hoc shift of bare dissipator}
This section will demonstrate that taking into consideration a shift and rotation when dealing with the dissipator allows the Extended Dicke model to relax to its energy minimum. It is stressed that this is not a formal derivation of dissipator (which is shown in section \ref{SectionDissipatorDressed}). Consequently, the dynamics of such a dissipative Dicke Model is not properly described by such a dissipator. However, the dissipator given here does demonstrate that if one takes into account the shift and rotation of the dissipator the system does relax to its energy minimum.

\begin{figure}
	\centering
	
	\includegraphics[width=0.4\textwidth]{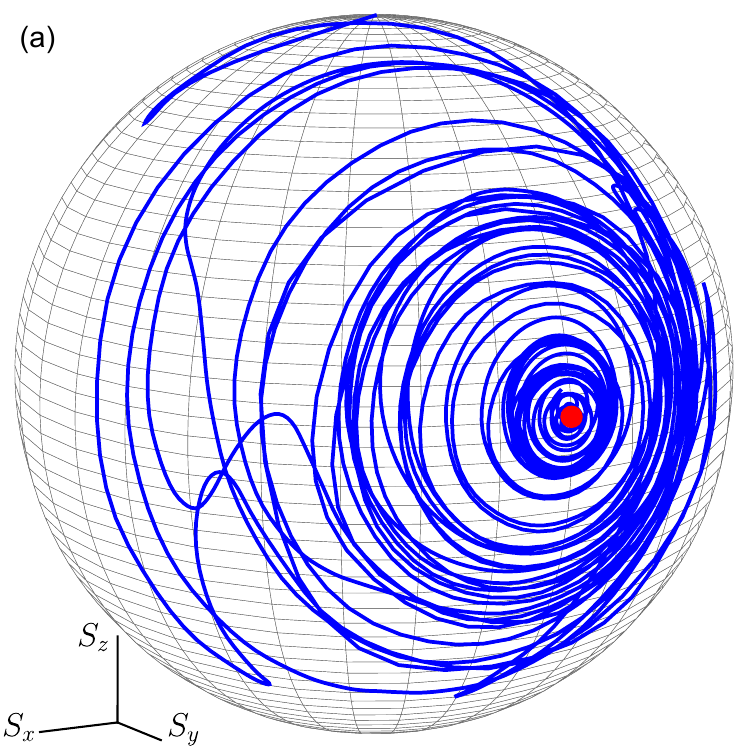}

	\includegraphics[width=0.4\textwidth]{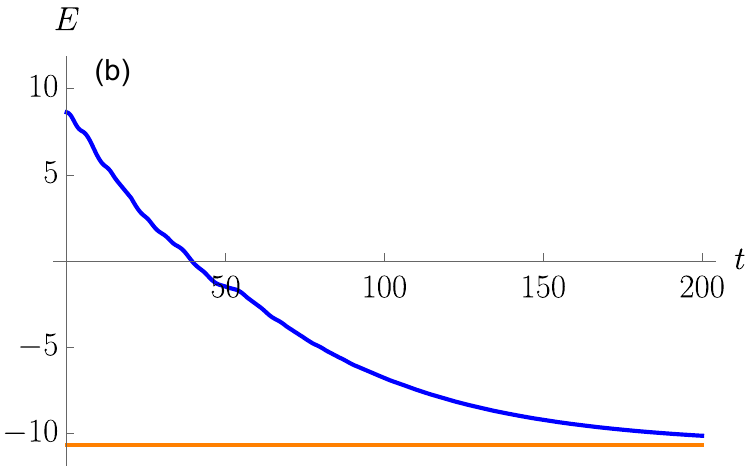}
	
	\caption{The semi-classical dynamics of the energy and superspin coordinates of the Extended Dicke model under influence of a rotated dissipator as described by Eq.(\ref{EqDissipatorRotated}). (Parameters: $\epsilon=-1, \omega=1,E_Z=0.2,g=0.46, \kappa_1=\kappa_2=0.02$). Here (Fig.\ref{FIG_asigma_motion_rotated}a) demonstrates the dynamics of the $S_{x,y,z}$-coordinates on a superspin sphere. The red points represent the coordinates of the minimum energy of the semi-classical approximation of the Extended Dicke model and the fixed points of the system. The second graph (Fig.\ref{FIG_asigma_motion_rotated}b) represents the dynamics of the energy corresponding to the same trajectory, the blue curve. The orange line represents the minimum energy. One can see that if dissipation is introduced to the Extended Dicke model by a rotated dissipator the system does relax to its minimum energy. This is in contrast with dynamics described by an unrotated dissipator as in Fig.\ref{FIG_asigma_motion} which do not relax to their minima in energy.}\label{FIG_asigma_motion_rotated} 
\end{figure}

To build a dissipator that does make the Extended Dicke model relax to its minimum, the $p$-coordinate needs to be shifted to $\hat{\tilde{p}} = \hat{p} - p_0$, $\hat{\tilde{q}} = \hat{q}$ and rotated towards the rotated frame (indicated by $\hat{J}_{x,y,z}$ with raising and lowering operators $\hat{J} = \hat{J}^\pm$) according to the transformation given by Eq.(\ref{EqAPPRotTrans}). The resulting operators are then $\mathcal{D}_{J^-}[\tilde{q},\tilde{p},J_{x,y,z}]$ and $\mathcal{D}_{\tilde{a}}[\tilde{q},\tilde{p},J_{x,y,z}]$. Using the linearity of the dissipator, one can then return into the unshifted and unrotated frame by Eq.(\ref{EqRotationDissipator}). The unitary part of the equations of motion remains the same as in Eq.(\ref{EqEOMBare}) with $\kappa_1 = \kappa_2 = 0$. The resulting dissipators have the following form (an example of a semi-classical ($\hat{O} \to O$) trajectory of such a system of equations is shown in Fig.\ref{FIG_asigma_motion_rotated}.):

\begin{gather}\label{EqDissipatorRotated}
    \begin{split}
	       \mathcal{D}[{\hat{q}}] & = -{\hat{q}}\\
	       \mathcal{D}[{\hat{p}}] & = -({\hat{p}}-p_0)\\
	       \mathcal{D}[{\hat{S}}_x] & = -{\hat{S}}_x\\
	       \mathcal{D}[{\hat{S}}_y] & = 2\sin(\theta)S - \left(1+\sin^2(\theta)\right){\hat{S}}_y - \sin(\theta)\cos(\theta){\hat{S}}_z\\
	       \mathcal{D}[{\hat{S}}_z] & = 2\cos(\theta)S - \sin(\theta)\cos(\theta){\hat{S}}_y - \left(1+\cos(\theta)\right){\hat{S}}_z.
    \end{split}
\end{gather}

In addition, it is noted that dissipation described by an 'ad hoc' shifted and rotated dissipator Eq.(\ref{EqDissipatorRotated}) has the same critical values of $g$ and $\epsilon$ as an undamped system.

\subsection{Dissipator built from the dressed creation/annihilation operators}\label{SectionDissipatorDressed}
It was shown in section \ref{SectionDissipatorBare} that the using a bare dissipator to describe the dynamics of the Exteneded Dicke model does not allow the system to relax to its ground state. More so, in section \ref{SectionShiftedHO} it was argued that the discrepancy of the fixed points of the 'bare' Lindblad equation with the coordinates of the minimum in energy of the Extended Dicke model are due to a shift and a rotation of the coordinates of the minimum in energy. In this section an overview will be given how the dissipator is built from the dressed or diagonalized lowering operators ($\hat{d}_{1,2}$) of $\hat{H}_{EDM}$ using the diagonalization presented in \cite{mukhin2018first}. 
A short overview of the prior mentioned diagonalizing will be given. In addition, the dissipators built from the operators $\hat{d}_{1,2}$ ($\mathcal{D}_{d_{1,2}}[\hat{O}]$) will be given. These dissipators will be used to derive the dissipators of the operators $\hat{q},\hat{p},\hat{S}_{x,y,z}$ ($\mathcal{D}_{d_{1,2}}[\hat{q},\hat{p},\hat{S}_{x,y,z}]$). For details, the reader is referred to Appendices \ref{APPDiagHamiltonian} (diagonalization) and \ref{APPDeriveMasterEq}, \ref{APPCoefficients} (derivation of the diagonalized dissipator).

\subsubsection{Derivation of Dissipator and diagonalization of Hamiltonian}
This section will calculate the dissipator for the extended Dicke model and its associated viscosity constant. This section will present the results of the calculations done by Beaudoin et al.\cite{beaudoin2011dissipation} and will calculate the associated matrix elements using a diagonalization from \cite{mukhin2018first}.

A formal derivation of the master equation / dissipator of system requires coupling the system to a bath often modelled as a collection of harmonic oscillators - a Caldeira-Leggett bath For coupling to the electromagnetic field such a coupling would look like $H_{SR} = \sum_{l}\alpha(\nu_l)\left(\hat{r}_l + \hat{r}_l^\dagger\right)\left(\hat{a} + \hat{a}^\dagger\right)$. Here $\hat{r}_l$, is the $l$-th mode of a bath of Harmonic oscillators, $H_R = \sum_l \nu_l \hat{r}^\dagger \hat{r}$, with the eigen frequences $\nu_l$, coupled to one of the system's operator $\hat{a}$, mediated by a coupling constant $\alpha(\nu_l)$. A similar coupling exists to a separate bath coupled to the spin/Holstein-Primakoff operators $H_{SR} = \sum_{l}\alpha'(\nu_l)\left(\hat{r}'_l + \hat{r'}_l^\dagger\right)\left(\hat{b} + \hat{b}^\dagger\right)$. We will derive the master equation for one dissipation channel. The other dissipation channel will be added at the end of the calculations for brevity and to prevent confusion. Such a master equation can be derived by setting up an equation of motion for the system combined with bath and tracing out the bath by assuming the bath will be at thermal equilibrium at all times. Assuming that the coupling to the bath is weak it is most convenient to derive such a master equation in the interaction representation. The interested reader is referred to Appendix \ref{APPDeriveMasterEq} for more details. Here the final result is presented for the master equation for a density matrix of a general Hamiltonian in the interaction representation:

\begin{gather}\label{EqMasterEqShiftTime}
    \begin{split}
        	\dot{\tilde{\rho}}(t)  = - & \sum_{j,k>j;m,n>m}C^*_{jk}C_{mn}\left(\ket{k}\bra{j}\ket{m}\bra{n}\tilde{\rho}(t) \right.\\
            &  \left.- \ket{m}\bra{n}\tilde{\rho}(t)\ket{k}\bra{j} \right)\times\\
		      & \times \left(\kappa_{nm} (1 + n_{nm}) + iJ_{nm} + iJ_{nm}' \right) + C.c.\\
		      - & \sum_{j,k>j;m,n>m}C^*_{jk}C_{mn}\left(\tilde{\rho}(t)\ket{m}\bra{n}\ket{k}\bra{j}\right. \\
           &   \left.- \ket{k}\bra{j}\tilde{\rho}(t)\ket{m}\bra{n}\right)\times\\
		      & \times \left(\kappa_{nm} n_{nm} + iJ_{nm}' \right) + C.c.
    \end{split}
\end{gather}

Here, $\tilde{\rho}(t)$ is the density matrix in the interaction representation. The states $\ket{n}$, are the n-th eigenstates of the Hamiltonian. The coefficients $\kappa_{nm}$, $n_{nm}$, $J_{nm}$ and $J'_{nm}$ are defined to be:
\begin{gather}\label{EqKappaJ}
    \begin{split}
        	\kappa_{nm} & = \pi g(\Delta_{nm})\alpha^2(\Delta_{nm})\\
	           n_{nm} & = n(\Delta_{nm},T)\\
	           J_{nm} & = \mathcal{P}\int_0^\infty d\nu \frac{g(\nu)\alpha^2(\nu)}{\Delta_{nm} - \nu}\\
	           J_{nm}'& = \mathcal{P}\int_0^\infty d\nu \frac{g(\nu)\alpha^2(\nu)n(\nu,T)}{\Delta_{nm} - \nu}\\
               C_{mn} & = \bra{m}\left(a+a^\dagger\right)\ket{n}.
    \end{split}
\end{gather}

Here $g(\nu)$ is the density of states at energy $\nu$. The states $\ket{m}$ and $\ket{n}$ are the eigenstates of the system. The Bose-Einstein distribution at temperature $T$ and energy $\Delta_{mn} = E_m-E_n$ is given by $n(\Delta_{mn},T)$. It is remarked that matrix elements similar to $C_{mn}$ exist for a bath connected to the Holstein-Primakoff-operators. 

Close attention needs to be put to the coefficients $C_{nm}$, which are given by $C_{mn} = \bra{m}\left(a+a^\dagger\right)\ket{n}$. This $C_{mn}$-coefficient allows to greatly simplify the master equation and in addition allows to determine the viscosity constant of the EDM. The study of $C_{mn}$ requires knowledge of the diagonalization employed\cite{mukhin2018first} to obtain the energy eigenstates.  For this reason a short outlining of this diagonalization is presented here. 
%%%%%%
The prior mentioned diagonalization assumes that the superspin is large ($S \gg 1$) (as would be the case in a quantum Dicke battery and other cases where the Extended Dicke Model would be applied). As a consequence of the large spin the tunneling between both of the semi-classical minima in energy can be assumed to be small. Due to the lack of tunneling, the symmetry of the Extended Dicke model is spontaneously broken and the diagonalization is performed in one of the semi-classical minima. The diagonalization rotates the spin in the $S_y$-$S_z$-plane by the quantum mechanical version of angle $\theta$ (see: Eq.(\ref{EqAPPAngelThetaQM})). In addition, the momentum is shifted by the quantum variant of $p^{SR}$. Consequently, the electromagnetic creation operators are also shifted: $\hat{a} = \hat{c} + i p_0 / \sqrt{2\omega}$ with $p_0 = \pm gS_y^{SR}$. The diagonalization is performed by using the Holstein-Primakoff transformation for the superspin for large S, followed by performing a Bogolyubov rotation by angle $\chi$ defined in Eq. (\ref{EqAPPFGKTan2Gamma}). Finally, the eigen energies $\varepsilon_{1,2}$ are determined and the Hamiltonian takes the form:

\begin{gather}
	\hat{H}_{EDM} = \varepsilon_1 \hat{d}^\dagger_1 \hat{d}_1 + \varepsilon_2 \hat{d}^\dagger_2 \hat{d}_2 + constant\\
    \varepsilon^2_{1,2} = \omega^2 + \frac{\omega E_Z}{\sqrt{-\varepsilon}\cos(\theta)}\frac{\cos(2\chi)\mp 1}{\sin(2\chi)},
\end{gather}

with the diagonalized lowering operators of the form: 
\begin{equation}
	\hat{d} = \alpha\hat{c} + \beta \hat{c}^\dagger + \gamma \hat{b} + \delta \hat{b}^\dagger .
\end{equation}

More specifically:

\begin{gather}\label{Eqdincb}
    \begin{split}
        \hat{d}_1 = & \frac{\cos(\chi)}{2}\left[\left(\sqrt{\frac{\omega}{\varepsilon_1}} - \sqrt{\frac{\varepsilon_1}{\omega}}\right)\hat{c}^\dagger + \left(\sqrt{\frac{\omega}{\varepsilon_1}} + \sqrt{\frac{\varepsilon_1}{\omega}}\right)\hat{c} \right]\\
        & + \frac{\sin(\chi)}{2}\left[\left(\sqrt{\frac{F}{\varepsilon_1}} - \sqrt{\frac{\varepsilon_1}{F}}\right)\hat{b}^\dagger + \left(\sqrt{\frac{F}{\varepsilon_1}} + \sqrt{\frac{\varepsilon_1}{F}}\right)\hat{b}\right]\\
        \hat{d}_2 = & - \frac{\sin(\chi)}{2}\left[\left(\sqrt{\frac{\omega}{\varepsilon_2}} - \sqrt{\frac{\varepsilon_2}{\omega}}\right)\hat{c}^\dagger + \left(\sqrt{\frac{\omega}{\varepsilon_2}} + \sqrt{\frac{\varepsilon_2}{\omega}}\right)\hat{c} \right]\\
        & + \frac{\cos(\chi)}{2}\left[\left(\sqrt{\frac{F}{\varepsilon_2}} - \sqrt{\frac{\varepsilon_2}{F}}\right)\hat{b}^\dagger + \left(\sqrt{\frac{F}{\varepsilon_2}} + \sqrt{\frac{\varepsilon_2}{F}}\right)\hat{b}\right].
    \end{split}
\end{gather}

Here $F=E_z/\cos(\theta)$.

The previous expressions (Eq.(\ref{Eqdincb})) allow to determine the coefficients $C_{mn}$. By letting the eigenstates act on the operator $a^\dagger + a$, it can be seen that only transitions to neighboring states are allowed. Taking into account, that, by design $n>m$,$k>j$, the $C_{mn}$-coefficients become:

\begin{gather}
    \begin{split}
        	C_{mn} = & \cos(\chi)\sqrt{\frac{\varepsilon_1}{\omega}}\sqrt{n_1}\delta_{m_1,n_1-1}\delta_{m_2,n_2}\\
            & + \sin(\chi)\sqrt{\frac{\varepsilon_2}{\omega}}\sqrt{n_2}\delta_{m_1,n_1}\delta_{m_2,n_2-1}\\
        	C^*_{jk} = & \cos(\chi)\sqrt{\frac{\varepsilon_1}{\omega}}\sqrt{j_1+1}\delta_{k_1,j_1+1}\delta_{k_2,j_2}\\
            & + \sin(\chi)\sqrt{\frac{\varepsilon_2}{\omega}}\sqrt{j_2+1}\delta_{k_1,j_1}\delta_{k_2,j_2+1}.
    \end{split}
\end{gather}

Attention is drawn to the product of matrix elements $C_{mn}C^*_{jk}$, which simplifies to a product of Kronecker deltas. We refer the reader for more details to Appendix \ref{APPCoefficients}, but it turns out that $C_{mn}C^*_{jk}$ only allows for either (Process 1) transitions between neighbouring elements of one of the two 'ladders' of Harmonic oscillator number states, or (Process 2) transitions from one of the two ladders of harmonic oscillator states to a neighbouring state on the other 'ladder'. All other transitions are completely suppressed. Additionally, the Process 1 transitions turn the exponent $\exp[i(\Delta_{mn}-\Delta_{jk})t]$ to unity. Process 2 transitions, however, oscillate rapidly. Making use of a Rotating wave approximation, the inter-ladder terms will be neglected in further calculations. Taking the sum over Kronecker deltas, identifying $\hat{d} = \sum_n \sqrt{n+1}\ket{n}\bra{n+1}$ as the lowering operator yields a great simplification. Considering, for brevity and clarity stake, first only one of the two dissipation channels and only $d_1$-mediated transition, the result for the density matrix in the interaction representation is given by:

\begin{gather}\label{EqLindbladDensityMatrix}
    \begin{split}
        	\dot{\tilde{\rho}} = & \cos^2(\chi)\frac{\varepsilon_1}{\omega}\left(iJ(\varepsilon_1)\left[\tilde{\rho}, \hat{d}_1^\dagger \hat{d}_1\right] + \kappa(\varepsilon_1)(n_B(\varepsilon_1)+1)\mathcal{D}_{d_1}[\tilde{\rho}]^\dagger\right)\\
             & + \cos^2(\chi)\frac{\varepsilon_1}{\omega}\left(\kappa(\varepsilon_1) n_B(\varepsilon_1) \mathcal{D}_{{d_1}^\dagger}[\tilde{\rho}]^\dagger \right),
    \end{split}
\end{gather}

where the conjugate of the dissipator is defined as $\mathcal{D}_a[\hat{O}]^\dagger = 2\hat{a}\hat{O}\hat{a}^\dagger - \{\hat{a}^\dagger \hat{a}, \hat{O}\}$. In addition, $n_B(\varepsilon_1) = n(\varepsilon_1,T)$, $\kappa(\varepsilon_1) = \pi g(\varepsilon_1) \alpha^2(\varepsilon_1)$, $J(\varepsilon_1) = \mathcal{P}\int d\nu g(\nu)\alpha^2(\nu)/(\varepsilon_1 - \nu)$.

The result in Eq. (\ref{EqLindbladDensityMatrix}) is remarkable, since it demonstrates that the only term contributing to the quantum viscosity coefficient that survives in zero temperature limit $T\rightarrow 0$ :  $\propto \kappa(\varepsilon_1)(n_B(\varepsilon_1)+1)$, with the Bose-Einstein function $n_B$ of harmonic oscillators in the thermal bath, indicating that virtual excitations of  harmonic oscillators in the thermal bath coupled to the polaritons of Dicke model give rise to effective viscosity in the $T\rightarrow 0$ limit. 

This fact sheds light on the quantum mechanism of viscosity when virtual excitations of the environment induced by the macroscopic system are taken into account. This fact adds  to the Landau criterium of superfluidity.  
 
The last term on the right hand side of Eq.(\ref{EqLindbladDensityMatrix}) disappears for $T=0$ since the number of Bose excitations $n_B(\varepsilon_1)$ goes to zero with temperature $T\rightarrow 0$ and hence will not contribute to viscosity. The first term, giving the Lamb shift, which has a form $i[H_{Lamb},O]$, gives rise to a modification of the unitary part of the equation of motion. It will not be considered here\cite{zueco2021photon}, because we focus merely on the non-unitary part of the equations of motion. What remains is to rotate back out of the interaction representation, adding the standard density matrix dynamics term back into the equation. The dissipator for an operator $\hat{O}$ in the Heisenberg picture and its associated dissipator will then simply be given by taking complex conjugate of the dissipator in the Schrodinger picture\cite{breuer2007opensystem}. The resulting equations of motion in the Heisenberg picture are then:
\begin{gather}
	 \dot{\hat{O}} = i[\hat{H},\hat{O}] +  \cos^2(\chi)\frac{\varepsilon_1}{\omega}\kappa(\varepsilon_1) \mathcal{D}_{d_1}[\hat{O}].
\end{gather}

Next, one recalls that only one dissipation channel was written down in Eq.(\ref{EqLindbladDensityMatrix}). Then, restoring the other channel, so that there would be one for photonic-bath coupling, and the other via spin-bath coupling, it is important to take into consideration that each dissipation channel gives both a transition via $d_1$ and via $d_2$. Hence, when all four contributions are considered in total, one obtains the final non-unitary part of the equations of motion (the 'dissipator') for an operator $\hat{O}$ with its associated viscocity constants:
\begin{gather}\label{EqDissipatorALL}
    \begin{split}
        	& \varepsilon_1 \left( \frac{\cos^2(\chi)\kappa_{a}(\varepsilon_1)}{\omega}+\frac{S \sin^2(\chi) \kappa_{S}(\varepsilon_1)}{2F} \right) \mathcal{D}_{d_1}[\hat{O}] \\
            & + \varepsilon_2 \left(\frac{\sin^2(\chi)\kappa_{a}(\varepsilon_2)}{\omega}+\frac{S\cos^2(\chi) \kappa_{S}(\varepsilon_2)}{2F}\right) \mathcal{D}_{d_2}[\hat{O}].
    \end{split}
\end{gather}

here the subscripts $a$ and $S$ refer to the two baths to which the system is coupled. Associatively the subscripts refer to the associated density of states ($g(\nu)$) and coupling constants ($\alpha(\nu)$). The subscript $a$ refers to the bath coupled to the electromagnetic field $\propto (a^\dagger + a)(r^\dagger +r)$. The subscript $S$ refers to a similar bath but coupled to spin-operators.

The expression Eq.(\ref{EqDissipatorALL}) can be abridged for the sake of argument to:
\begin{gather}
	\kappa_1 \mathcal{D}_{d_1}[\hat{O}] + \kappa_1 \mathcal{D}_{d_2}[\hat{O}].
\end{gather}

The diagonalized raising and lowering operators from Eq.\ref{EqAPPOpscbIndd} can then be used to calculate the dissipator by assuming the Extended Dicke model is coupled to a Caldeira-Leggett bath of harmonic oscillators. These derivations (presented in Appendix \ref{APPDeriveMasterEq}) yield the following dissipators:
\begin{align}\label{EqDissipatorDressed}
	\mathcal{D}_{d_{1,2}}[\hat{O}] = 2 \hat{d}_{1,2}^\dagger \hat{O} \hat{d}_{1,2} - \{ \hat{d}_{1,2}^\dagger \hat{d}_{1,2},\hat{O} \}.
\end{align}

\subsubsection{Dissipators of $p,q,S_{x,y,z}$}\label{SectionDissipatorOfpqS}
The dissipators $\mathcal{D}_{d_{1,2}}[\hat{O}]$ allow to determine the non-unitary part of the equations of motion for the operators $\hat{q},\hat{p},\hat{S}_{x,y,z}$. In order to do this, first the dissipators of the shifted photonic destruction operator ($\hat{c}$), the rotated Holstein-Primakoff destruction and number operators ($\hat{b}$ and $\hat{b}^\dagger b$) will be calculated. Afterwards dissipators of the shifted and rotated photonic and spin operators ($\hat{\tilde{q}},\hat{\tilde{p}},\hat{J}_{x,y,z}$) are determined. Finally, having obtained $\mathcal{D}_{d_{1,2}}[\hat{\tilde{q}},\hat{\tilde{p}},\hat{J}_{x,y,z}]$, these operators will be shifted and rotated in the $q,p,S_{x,y,z}$-frame.

One starts by noticing that the shifted position and momentum operators $\hat{\tilde{q}} = \sqrt{\frac{1}{2\omega}}(\hat{c}^\dagger + \hat{c})$, $\hat{\tilde{p}} = i\sqrt{\frac{\omega}{2}}(\hat{c}^\dagger -\hat{c})$ and the rotated spin operators $\hat{J}_x = \sqrt{\frac{S}{2}}\left(\hat{b}^\dagger + \hat{b} \right)$ are linear combinations of the operators $\hat{c}$ and $\hat{b}$. Then, one can use the linearity of the dissipator ($\mathcal{D}[a\hat{A} + b \hat{B}] = a\mathcal{D}[\hat{A}] +  b\mathcal{D}[\hat{B}]$) and by using the property $\mathcal{D}[O^\dagger] = \mathcal{D}[O]^\dagger$ to determine the dissipators for the $\hat{\tilde{p}}, \hat{\tilde{q}}, \hat{J}_x, \hat{J}_y$. The dissipator for $\hat{J}_z = S - \hat{b}^\dagger \hat{b}$ needs to be calculated separately. After this procedure one finds the dissipator in the shifted and rotated frame given in Eq.(\ref{EqDissipatorRotatedDressed}).

The next step is to transform these shifted and rotated dissipators back into the normal frame. Using the shift and rotation performed to move into the rotated frame Eq.(\ref{EqAPPRotTrans}), one can again use the linearity of the dissipator to find the various unrotated and unshifted dissipators in terms of the shifted, rotated ones (see: Eq.(\ref{EqRotationDissipator})). Finally, one is able to find the dissipators in the normal frame:
\begin{gather}\label{EqDissipatorNormalDressed}
    \begin{split}
            \mathcal{D}_{d_{1,2}}[\hat{q}] = & - A_{1,2} \hat{q} + B_{1,2} \hat{S}_x\\
        	\mathcal{D}_{d_{1,2}}[\hat{p}] = & - A_{1,2} \left(\hat{p}-p_0\right) + C_{1,2} \cos(\theta)\hat{S}_y - C_{1,2}\sin(\theta)\hat{S}_z\\
        	\mathcal{D}_{d_{1,2}}[\hat{S}_x] = & - D_{1,2} \hat{S}_x + E_{1,2} \hat{q}\\
        	\mathcal{D}_{d_{1,2}}[\hat{S}_y] = & F_{1,2}\cos(\theta)(\hat{p}-p_0) + C_{1,2}\sin(\theta)\hat{q}\hat{S}_x +\\ 
      		 &2D_{1,2}\sin(\theta)S - \left[D_{1,2}(1+\sin^2(\theta)) -\right.\\
		& \left.- \frac{B_{1,2}\sin(2\theta)}{2}(\hat{p}-p_0)\right]\hat{S}_y+\\
        		& + \left[-B_{1,2}\sin^2(\theta)(\hat{p}-p_0) - \frac{D_{1,2}}{2}\sin(2\theta)\right]\hat{S}_z\\
        	\mathcal{D}_{d_{1,2}}[\hat{S}_z] = & 2D_{1,2}\cos(\theta)S - F_{1,2}\sin(\theta)(\hat{p}-p_0) +  \\
        		& C_{1,2}\cos(\theta)\hat{q}\hat{S}_x - \left[\frac{D_{1,2}}{2}\sin(2\theta) - \right.\\
		&\left. -B_{1,2}\cos^2(\theta)(\hat{p}-p_0))\right]\hat{S}_y+\\
        		+ &\left[-\frac{B_{1,2}}{2}\sin(2\theta)(\hat{p}-p_0) - D_{1,2}\left(1+\cos^2(\theta)\right)\right]\hat{S}_z
    \end{split}
\end{gather}

where $-p_0=\pm gS_{y}^{SR}$, $\sin(\theta) = S_y^{SR}/S$, $\cos(\theta) = S_z^{SR}/S$. The other constants are defined as:
\begin{gather}
    \begin{split}
        & 2A_1 = 2D_2 = \cos^2(\chi)\\
        & 2A_2 = 2D_1 = \sin^2(\chi)\\
        & 2B_{1,2} = \mp \sin(\chi)\cos(\chi)\frac{1}{\omega}\sqrt{\frac{F}{S}}\\
        & 2C_{1,2} = \mp \sin(\chi)\cos(\chi) \frac{\omega}{\sqrt{SF}}\\
        & 2E_{1,2} = \mp \sin(\chi)\cos(\chi) \omega \sqrt{\frac{S}{F}}\\
        & 2F_{1,2} = \mp \sin(\chi)\cos(\chi) \frac{\sqrt{SF}}{\omega}
    \end{split}
\end{gather}

The remarkable thing is that, despite arising from a different creation/annihilation operator, both the $\mathcal{D}_{d_1}$- and the  $\mathcal{D}_{d_2}$-dissipators are exactly the same except for two things: 1. $A_1$ and $D_2$ are cosines, their associated parameters from the other dissipative channel give sines. 2. All other parameters of the dissipation channel of $d_1$ are opposite in sign (but equal in norm) to their associated parameters of the $d_1$ dissipation channel. Refering to Eq.(\ref{EqDissipatorALL}) and Eq.(\ref{EqDissipatorNormalDressed}), one can see that, except for the terms $A_{1,2}$ and $D_{1,2}$ all other terms of the dissipators work in opposite directions due to the opposite signs. The only terms that reinforce each other (belonging to $A_{1,2}$ and $D_{1,2}$ give rise to the same form of the 'ad hoc shifted dissipator' of Eq.(\ref{EqDissipatorRotated}). Indeed, the terms that can in principle have both negative or positive signs cannot contribution to energy loss, otherwise they could pump energy into the system. The only that are responsible for energy loss are those that always have the same (negative) sign, i.e. those terms corresponding to the 'ad hoc' shifted/rotated dissipator.

The dissipative equations of motion are found by taking unitary Heisenberg equation of motion (found from Eq.(\ref{EqEOMBare}) by setting $\kappa_1,\kappa_2 = 0$) and adding to them the 'dressed' dissipators from Eq.(\ref{EqDissipatorNormalDressed}) ($\dot{\hat{O}} = i[\hat{H},\hat{O}]+\kappa \mathcal{D}_L[\hat{O}]$). These quantum equations of motion can be converted into semi-classical equations of motion by changing the quantum operators to c-numbers ($\hat{O} \to O$). The fixed points of the semi-classical equation of motion can be found by setting the time derivatives equal to zero ($\dot{O} = 0$). By plugging in the coordinates of the minimum energy into the 'dressed' dissipative equations of motion one can establish that these 'dressed' equation of motion indeed have a fixed point at minimum energy. In addition, the dissipators in Eq.(\ref{EqDissipatorNormalDressed}) do not shift critical values for $g$ and $\epsilon$ like a 'bare' dissipators do (see: Eq.(\ref{EqEOMBare})).

Finally, it must be noted that the diagonalization in \cite{mukhin2018first} assumes that the superspin is close to its maximum value ($N/2$). The assumption  allows to approximate the Holstein-Primakoff transformation as $\hat{S}^- \approx \sqrt{S/2}\hat{b}$, which is only valid if the system remains close to maximum superspin. Hence, the diagonalization and consequently the dissipators and equations of motion derived from it are only valid in the proximity of minimum in energy.

\section{Discussion}

{A derivation was presented of a quantum master equation for the extended Dicke model valid in the USC-regime / superradiant phase. The master equation was presented in the Heisenberg picture for the Dicke model operators $q$, $p$ and $S_{x,y,z}$. Specifically, the dissipator derived is valid near the minimum in energy in the non-tunneling, spontaneous symmetry broken state of the Dicke model at a large number $N$ of Two-level systems. An associated expression for the viscosity of such a system has also been found.}

{The presence of a large amount of  two-level systems (pseudo spins) in the Dicke Model allows for a semi-classical approximation to many quantum mechanical phenomena that are comprised by the model. In the present study, dissipation was studied by deriving quantum mechanical and semi-classical equation of motion in the presence of dissipation using Lindblad type dissipators. It was shown, that building a dissipator by using the 'bare' creation/anihilation operators can pump a macroscopic amount of energy into the system. Indeed in the USC-regime/superradiant phase, the system no longer relaxes to the energy minimum under the influence of such a 'bare' dissipation. It was then argued that upon a c-number shift and a rotation of the operators under the superradiant transition in the dissipator operator, the  system would again relax to its ground state. In addition, a full derivation of a quantum mechanical master equation has been performed using diagonalized Hamiltonian in the superradiant set of the eigenstates, yielding the dissipators in the Heisenberg picture of the $q$-, $p$- and $S_{z,y,z}$-operators. It was shown that under the influence of such a master equation, the system does indeed relax to its semi-classical energy minimum. It was shown, that such a master equation corresponds exactly to the shifted / rotated dissipator for the terms responsible for energy loss. It was argued that the classical reason for the failure of the 'bare' dissipator to let the system relax to its energy minimum is due to a failure to incorporate the shift of its p-coordinate corresponding to the new energy minimum and a rotation of its $S_y$- and $S_z$-coordinates when the system goes into the USC-regime / superradiant state.}

Considering the derivation of Eq.(\ref{EqDissipatorNormalDressed}), an important assumption made is the explicit symmetry breaking already present in the diagonalization presented in \cite{mukhin2018first}. The diagnolization, in the first steps of the derivation, shifts the $p$-operator and rotates the spin-operators to one of the minima in energy. Of course, this does not account for tunneling. Tunneling will, of course, be present, but it will be negligible: in the presence of a macroscopically large spin such an assumption is sound for the minima are well separated. In addition, because the diagonalization in Ref.\cite{mukhin2018first} makes use of the Holstein-Primakoff transformation and assumes that superspin is close to $S=N/2$, the diagonalization is valid only close to the minimum in energy. Consequently, validity of the dissipators and consequent equations of motions derived in that section is also based on the proximity of minimum in energy due to making use of the diagonalization results in \cite{mukhin2018first}.

\section{Conclusion}

Quantum dissipation described by Lindblad-type master equations has been studied in the USC-regime using a semi-classical approach. It has been argued, by studying the extended Dicke model allowing for a classical approach at $N\gg1$ spins, that the classical reason why Lindblad-type master equations, built from 'bare' creation/annihilation operators, fails to let systems relax to its ground state is due to them being unable to account for a shift in the coordinates of the minimum in energy. Hence, such 'bare' Lindblad equations 'drag' the system away from its minimum and consequently 'pump' energy into the system. Besides, a quantum and semi-classical dissipator have been found by deriving a quantum master equation using a Caldeira-Leggett type bath, which does allow the system to relax to its ground state. The resulting equation of motion allows to describe the dynamics of extended Dicke model semi-classically in the spontaneously symmetry broken, non-tunneling regime at low energies. The effective viscosity in the semi-classical dynamics equations of motion of the superradiant phase remains nonzero at T=0 due to virtual excitations of  harmonic oscillators in the thermal bath coupled to the polaritons of the Dicke model. 

\section*{Acknowledgements}
This work was supported by the Ministry of Science and Higher Education of the Russian Federation in the framework of the Strategic Academic Leadership program “Priority 2030” (MISIS Strategic Technology Project ‘Quantum Internet’).

\appendix

\section{Diagonalization of the $H_{EDM}$ and associated parameters simplifications in the large spin regime}\label{APPDiagHamiltonian}
This article has made extensive use of the diagonalization of the Extended Dicke model preformed in \cite{mukhin2018first}. In this section, a summary is given of the diagonalization and expressions used in the diagonalization are defined. In addition, these expressions will be examined for large values of the total spin $S$.\\
In \cite{mukhin2018first} the diagonalization of the Extended Dicke model has been performed as follows. The photonic creation/annihilation operators $\hat{a}$, are shifted ($\hat{a} = \hat{c} + i\alpha$) in order to accommodate the shift of the photonic momentum in its ground state. Similarly, the spin-operators $S_{x,y,z}$ are rotated by $\theta$ to accommodate a non-zero value of the $S_y$-operator at zero energy. The transformation is given by:
\begin{gather}\label{EqAPPRotTrans}
    \begin{split}
    	\hat{S}_x & = \hat{J}_x\\
    	\hat{S}_y & = \hat{J}_z \sin(\theta) + \hat{J}_y \cos(\theta)\\
    	\hat{S}_z & = \hat{J}_z \cos(\theta) - \hat{J}_y \sin(\theta).
    \end{split}
\end{gather}

As used in \cite{mukhin2018first}, a Holstein-Primakoff transformation of the spin operators results in $\hat{S}_- \to \sqrt{\frac{S}{2}}\hat{b} $ at large values of the total spin $S$. The Hamiltonian is then diagonalized in the form ($\varepsilon_{1,2}$ and $G$ to be defined shortly):
\begin{gather}\label{EqHamiltonianDiag}
	\hat{H} = \varepsilon_1 \hat{d}_1^\dagger \hat{d}_1 + \varepsilon_2 \hat{d}_2^\dagger \hat{d}_2 + \frac{\varepsilon_1 + \varepsilon_2}{2} + G.
\end{gather}

In what follows in this section, the expression will be defined and examined at large $S$. First, look at the angle $\theta$. It must be noted that, in order to examine the superradiant state, $\theta$ needs to have a finite value.
\begin{gather}\label{EqAPPAngelThetaQM}
    \begin{split}
        \cos(\theta) &  = \frac{2E_Z/g^2}{1+\varepsilon - 2\varepsilon \left(S - \langle b^\dagger b \rangle\right)}\\
		& =^{S \gg 1} -\frac{E_Z}{\varepsilon g^2 S}.
    \end{split}
\end{gather}
In correspondence with equation Eq.(\ref{EqAngelTheta}).

In anticipation of the definitions for the energies $\varepsilon_{1,2}$, the quantities $F$, $G$, $K$ and a second Bogolyubov angle $\chi$ are defined to be:
\begin{gather}\label{EqAPPFGKTan2Gamma}
    \begin{split}
        F = & \frac{E_Z}{\cos(\theta)}\\
        G = & - \frac{E_Z S}{2}\left(\frac{1 + \cos^2(\theta)}{\cos(\theta)}\right)\\
	    K = & (1+\varepsilon)g^2/2\\
	    \tan(2\chi) = & 2\frac{\frac{E_Z}{\sqrt{-\varepsilon}\omega \cos(\theta)}}{\frac{E_Z^2}{-\varepsilon\omega^2 \cos^2(\theta)} - 1}.
    \end{split}
\end{gather}

Next, one is able to define the energies $\varepsilon_{1,2}$ in terms of the previous expression. Diagonalization of the Extended Dicke model gives the following eigenvalues:
\begin{gather}\label{EqAPPEnergyQuantum}
    \begin{split}
        	2\varepsilon^2_{1,2} = & \omega^2 + F^2 + 2SKF \mp \left(F^2 + 2SKF - \omega^2\right)\cos(2\chi) \\
            & \mp 2g\omega\sqrt{SF}\sin(2\chi).
    \end{split}
\end{gather}

The expressions for $\varepsilon_{1,2}$ can be simplified under the assumptions made earlier. After plugging Eq.(\ref{EqAPPFGKTan2Gamma}) into Eq.(\ref{EqAPPEnergyQuantum}), one obtains:

\begin{gather}\label{EqAPEnergyClassical}
	\varepsilon^2_{1,2} = \omega^2 + \frac{\omega E_Z}{\sqrt{-\varepsilon}\cos(\theta)}\frac{\cos(2\chi)\mp 1}{\sin(2\chi)}.
\end{gather}

The operators $\hat{c}, \hat{b}$, can be expressed in the diagonalized creation and annihilation operators as:
\begin{gather}\label{EqAPPOpscbIndd}
    \begin{split}
    	\hat{c} = & \frac{\cos(\chi)}{2}\left[\left(\sqrt{\frac{\varepsilon_1}{\omega}} - \sqrt{\frac{\omega}{\varepsilon_1}}\right)\hat{d}^\dagger_1 + \left(\sqrt{\frac{\varepsilon_1}{\omega}} + \sqrt{\frac{\omega}{\varepsilon_1}}\right)\hat{d}_1 \right]\\
    	& + \frac{\sin(\chi)}{2}\left[\left(\sqrt{\frac{\varepsilon_2}{\omega}} - \sqrt{\frac{\omega}{\varepsilon_2}}\right)\hat{d}^\dagger_2 + \left(\sqrt{\frac{\varepsilon_2}{\omega}} + \sqrt{\frac{\omega}{\varepsilon_2}}\right)\hat{d}_2 \right]\\
    	\hat{b} = & - \frac{\sin(\chi)}{2}\left[\left(\sqrt{\frac{\varepsilon_1}{F}} - \sqrt{\frac{F}{\varepsilon_1}}\right)\hat{d}^\dagger_1 + \left(\sqrt{\frac{\varepsilon_1}{F}} + \sqrt{\frac{F}{\varepsilon_1}}\right)\hat{d}_1 \right]\\
    	& + \frac{\cos(\chi)}{2}\left[\left(\sqrt{\frac{\varepsilon_2}{F}} - \sqrt{\frac{F}{\varepsilon_2}}\right)\hat{d}^\dagger_2 + \left(\sqrt{\frac{\varepsilon_2}{F}} + \sqrt{\frac{F}{\varepsilon_2}}\right)\hat{d}_2 \right].
    \end{split}
\end{gather}

Similarly, it is possible to express the operators $\hat{d}_{1,2}$ in terms of $\hat{c}$ and $\hat{b}$. Such an inversion gives the following form for the operators $\hat{d}_{1,2} = \alpha_{1,2} \hat{c} + \beta_{1,2} \hat{c}^\dagger + \gamma_{1,2}b + \delta_{1,2}b^\dagger$. The constants $\alpha_{1,2}$, $\beta_{1,2}$, $\gamma_{1,2}$, $\delta_{1,2}$ can be found from:

\begin{gather}\label{EqAPPOpsddIncb}
    \begin{split}
        \hat{d}_1 = & \frac{\cos(\chi)}{2}\left[\left(\sqrt{\frac{\omega}{\varepsilon_1}} - \sqrt{\frac{\varepsilon_1}{\omega}}\right)\hat{c}^\dagger + \left(\sqrt{\frac{\omega}{\varepsilon_1}} + \sqrt{\frac{\varepsilon_1}{\omega}}\right)\hat{c} \right]\\
        & + \frac{\sin(\chi)}{2}\left[\left(\sqrt{\frac{F}{\varepsilon_1}} - \sqrt{\frac{\varepsilon_1}{F}}\right)\hat{b}^\dagger + \left(\sqrt{\frac{F}{\varepsilon_1}} + \sqrt{\frac{\varepsilon_1}{F}}\right)\hat{b}\right]\\
        \hat{d}_2 = & - \frac{\sin(\chi)}{2}\left[\left(\sqrt{\frac{\omega}{\varepsilon_2}} - \sqrt{\frac{\varepsilon_2}{\omega}}\right)\hat{c}^\dagger + \left(\sqrt{\frac{\omega}{\varepsilon_2}} + \sqrt{\frac{\varepsilon_2}{\omega}}\right)\hat{c} \right]\\
        & + \frac{\cos(\chi)}{2}\left[\left(\sqrt{\frac{F}{\varepsilon_2}} - \sqrt{\frac{\varepsilon_2}{F}}\right)\hat{b}^\dagger + \left(\sqrt{\frac{F}{\varepsilon_2}} + \sqrt{\frac{\varepsilon_2}{F}}\right)\hat{b}\right]\\
    \end{split}
\end{gather}

\section{Derivation of the Dissipator}\label{APPDeriveMasterEq}
This appendix shortly outlines the derivation of the dissipator by using a Caldeira-Leggett bath of harmonic oscillators. This derivation is based on the description of dissipation in the strong coupling regime in the Rabi-model previously done by Beaudoin et al. \cite{beaudoin2011dissipation}. For brevity it has been assumed that the dissipation mediated by the photonic-creation/annihilation operators communicates with a completely different bath than the dissipation mediated by the spin creation/annihilation operators. For the sake of  clarity, the equation of motion/dissipator for a density matrix is derived first. This dissipator can then easily be transformed to describe non-unitary motion for operators as was done in the remainder of this article.

Assumed is that the system with bare Hamiltonian ($\hat{H}_S = \sum_j E_j \ket{j}\bra{j}$) couples weakly to a reservoir/bath ($\hat{H}_R = \sum_l \nu_l \hat{r}_l^\dagger \hat{r}_l$) via a system-bath term mediated by the photonic creation/annihilation operators ($\hat{a}$):
\begin{equation}
	\hat{H}_{SR} = \sum_{l}\alpha(\nu_l)\left(\hat{r}_l + \hat{r}_l^\dagger\right)\left(\hat{a} + \hat{a}^\dagger\right),
\end{equation}

where $\alpha_l$ is the coupling constant to the $l$-bath mode.

Preparing to move into the interaction representation, this expression can be rewritten by noting that $\hat{R} = \sum_l\alpha(\nu_l) \hat{r}_l$ and expressing the operator $\hat{a}+\hat{a}^\dagger$ in terms of energy eigenstates of $H_s$. Defining $C_{jk} = \bra{j}\left(a+a^\dagger\right)\ket{k}$, and a lowering operator $\hat{s} = \sum_{j,k>j}C_{jk}\ket{j}\bra{k}$, gives:
\begin{equation}
	\hat{H}_{SR} = \left(\hat{R}+\hat{R}^\dagger\right)\left(\hat{s}+\hat{s}^\dagger\right).
\end{equation}

The diagonal matrix elements equal zero ( $C_{jj}=0$) because in the Extended Dicke model Hamiltonian parity is a good quantum number and the operator $\hat{c}+\hat{c}^\dagger$ changes parity. Making use of the standard methods to obtain a Born-Markov master equation in the interaction representation \cite{carmichael1991opensystem}, one is able to obtain an equation of motion for the density matrix of the system ($\hat{\rho}$). In short, this is done by setting up a Von Neumann equation of motion for the density matrix of system+bath in the interaction representation and tracing by assuming it is in thermal equilibrium. The procedure leads to an equation for the time evolution density matrix of the system ($\hat{\rho}$) in the interaction representation (here indicated by tildes).
\begin{equation}
	\dot{\hat{\tilde{\rho}}}(t) = - \int_0^t dt' \tr_R \{\left[\hat{\tilde{H}}_{SR}(t),\left[ \hat{\tilde{H}}_{SR}(t') , \hat{\tilde{\rho}}(t) \hat{R}_0\right]\right]\}.
\end{equation}

Opening up the commutators and simplifying the resulting expression by noting that $\langle \hat{r}_i^\dagger \hat{r}_j^\dagger \rangle = \langle \hat{r}_i \hat{r}_j \rangle = 0$ yields that only terms containing both a $\hat{R}^\dagger$ and $\hat{R}$ survive the tracing. Defining $t-t' = \tau$ and $n(\nu_l,T) = (\exp(\nu_l/T)+1)^{-1}$, one can write:
\begin{gather}
    \begin{split}
        &\langle \hat{R}^\dagger(t)\hat{R}(t') \rangle = \sum_{l}\alpha(\nu_l)^2 n(\nu_l, T)e^{i\nu_l (t-t')}\\
	&\langle \hat{R}(t)\hat{R}^\dagger(t') \rangle = \sum_{l}\alpha(\nu_l)^2 \left(1+n(\nu_l, T)\right)e^{-i\nu_l (t-t')}.
    \end{split}
\end{gather}

Plugging the previous two equations into the Master equation (and defining $\Delta_{mn} = E_m - E_n$), one obtains
\begin{gather}
    \begin{split}
        	\dot{\hat{\tilde{\rho}}}(t)  = - & \int_0^t dt'\sum_{j,k>j;m,n>m}C^*_{jk}C_{mn}\left(\ket{k}\bra{j}\ket{m}\bra{n}\hat{\tilde{\rho}}(t')\right.\\
           & \left. - \ket{m}\bra{n}\hat{\tilde{\rho}}(t')\ket{k}\bra{j} \right)\times\\
		& \times \sum_l \alpha_l^2(1+n(\nu_l,T))e^{i (\Delta_{mn}-\Delta_{jk})t + i(\Delta_{mn}- \nu_l) \tau}\\
        & + C.c.\\
		- & \int_0^t dt'\sum_{j,k>j;m,n>m}C^*_{jk}C_{mn}\left(\hat{\tilde{\rho}}(t')\ket{m}\bra{n}\ket{k}\bra{j}\right.\\
        & \left. - \ket{k}\bra{j}\hat{\tilde{\rho}}(t')\ket{m}\bra{n}\right)\times\\
		& \times \sum_l \alpha_l^2 n(\nu_l,T)e^{i (\Delta_{mn}-\Delta_{jk})t + i(\Delta_{mn}- \nu_l) \tau} + C.c.
    \end{split}
\end{gather}

In what follows  (using standard methods \cite{carmichael1991opensystem}) the integration variable is switched to $\tau$. Assuming that the integration over $\tau$ is dominated by times that are much shorter than the time scale for the evolution of $\hat{\tilde{\rho}}(t)$, $\hat{\tilde{\rho}}(t-\tau)$ is approximated by $\hat{\tilde{\rho}}(t)$ (the Markovian approximation). This step allows to take the density matrix outside of the integral. Under the Markovian assumption, one can safely let the boundary of the integral go to infinity ($t \to \infty$). In addition, the sum over all possible bath modes can be approximated as an integral over $\nu$ with mode density function $g(\nu)$. After making use of the Sokhotski-Plemelj identity, one obtains:
\begin{gather}\label{EqAPPMasterEqShiftTime}
    \begin{split}
        	\dot{\hat{\tilde{\rho}}}(t)  = - & \sum_{j,k>j;m,n>m}C^*_{jk}C_{mn}\left(\ket{k}\bra{j}\ket{m}\bra{n}\hat{\tilde{\rho}}(t) \right.\\
            &  \left.- \ket{m}\bra{n}\hat{\tilde{\rho}}(t)\ket{k}\bra{j} \right)\times\\
		      & \times \left(\kappa_{nm} (1 + n_{nm}) + iJ_{nm} + iJ_{nm}' \right) + C.c.\\
		      - & \sum_{j,k>j;m,n>m}C^*_{jk}C_{mn}\left(\hat{\tilde{\rho}}(t)\ket{m}\bra{n}\ket{k}\bra{j}\right. \\
           &   \left.- \ket{k}\bra{j}\hat{\tilde{\rho}}(t)\ket{m}\bra{n}\right)\times\\
		      & \times \left(\kappa_{nm} n_{nm} + iJ_{nm}' \right) + C.c.
    \end{split}
\end{gather}

where: 
\begin{gather}\label{EqAPPKappaJ}
    \begin{split}
        	\kappa_{nm} & = \pi g(\Delta_{nm})\alpha^2(\Delta_{nm})\\
	           n_{nm} & = n(\Delta_{nm},T)\\
	           J_{nm} & = \mathcal{P}\int_0^\infty d\nu \frac{g(\nu)\alpha^2(\nu)}{\Delta_{nm} - \nu}\\
	           J_{nm}'& = \mathcal{P}\int_0^\infty d\nu \frac{g(\nu)\alpha^2(\nu)n(\nu,T)}{\Delta_{nm} - \nu}.
    \end{split}
\end{gather}

\section{Calculation of the $C_{mn}$ coefficients.}\label{APPCoefficients}
This section will focus on calculating the coefficients $C_{mn}$ required to determine the dissipator. In addition, the product $C_{mn}C^*_{jk}$ will be examined. A similar approach will be used as was pioneerd in ref.\cite{zueco2021photon}. Starting by recalling that the $C_{mn}$-matrix elements are calculated by letting the eigenstates of the Extended Dicke model ($\ket{m},\ket{n}$) act on the operators Bosonic $\hat{a}^\dagger + \hat{a}$ and and Spin $\hat{S}^+ + \hat{S}^-$ operators. It can be easily seen in Eq.(\ref{EqAPPRotTrans}) that the shift of the bosonic operator, respectively, rotation of the spin operators do not change the form of the shifted/rotated operators, giving: $\hat{c}+\hat{c}^\dagger$ and $\hat{J}^- + \hat{J}^+$. In terms of the diagonalized operators $\hat{d}_{1,2}$, the sum of the photonic and sum of the spin operators can be expressed as:
\begin{gather}
    \begin{split}
    	\hat{a}^\dagger + \hat{a} = & + \cos(\chi)\sqrt{\frac{\varepsilon_1}{\omega}}\left(\hat{d}_1^\dagger + \hat{d}_1\right)\\
         &+ \sin(\chi)\sqrt{\frac{\varepsilon_2}{\omega}}\left(\hat{d}_2^\dagger + \hat{d}_2\right)\\
    	\hat{S}^+ + \hat{S}^- = & -\sin(\chi)\sqrt{\frac{S\varepsilon_1}{2F}}\left(\hat{d}_1^\dagger + \hat{d}_1\right)\\
         & + \cos(\chi)\sqrt{\frac{S \varepsilon_2}{2 F}}\left(\hat{d}_2^\dagger + \hat{d}_2\right).
    \end{split}
\end{gather}

Using the above operator, one can calculate the matrix element $C_{mn}$. To be more specific, however, it needs to be pointed out that the states $\ket{m}, \ket{n}$ can be written in terms of the occupation numbers of the two Harmonic Oscillators obtained after diagonalization (states created by $\hat{d}_1^\dagger$ and $\hat{d}_2^\dagger$). In terms of the number states of the harmonic oscillator, one can write $\ket{m}= \ket{m_1,m_2}$, $\ket{n}= \ket{n_1,n_2}$. Letting these states act on the operator $\hat{a}^\dagger + \hat{a}$ and keeping in mind that, by design $n>m$, $k>j$, one obtains:
\begin{gather}
    \begin{split}
        	C_{mn} = & \cos(\chi)\sqrt{\frac{\varepsilon_1}{\omega}}\sqrt{n_1}\delta_{m_1,n_1-1}\delta_{m_2,n_2}\\
            & + \sin(\chi)\sqrt{\frac{\varepsilon_2}{\omega}}\sqrt{n_2}\delta_{m_1,n_1}\delta_{m_2,n_2-1}\\
        	C^*_{jk} = & \cos(\chi)\sqrt{\frac{\varepsilon_1}{\omega}}\sqrt{j_1+1}\delta_{k_1,j_1+1}\delta_{k_2,j_2}\\
            & + \sin(\chi)\sqrt{\frac{\varepsilon_2}{\omega}}\sqrt{j_2+1}\delta_{k_1,j_1}\delta_{k_2,j_2+1}.
    \end{split}
\end{gather}
A similar relation holds for the matrix elements $\bra{m}(\hat{S}^- + \hat{S}^+)\ket{n}$ of the spin raising/lowering operators. 

Next, attention is focused on the product of the matrix elements $C_{mn}C^*_{jk}$, as is occurring in the derivation of the master equation. Up to some factor (indicated by $(\cdots)$), the expressions look like:
\begin{gather}\label{EqAPP_CmnCjk_Delta}
    \begin{split}
    	C_{mn}C^*_{jk} = & (\cdots)\sqrt{n_1}\sqrt{j_1+1}\delta_{m_1,n_1-1}\delta_{m_2,n_2}\delta_{k_1,j_1+1}\delta_{k_2,j_2} \\
      & + (\cdots)\sqrt{n_1}\sqrt{j_2+1}\delta_{m_1,n_1-1}\delta_{m_2,n_2}\delta_{k_1,j_1}\delta_{k_2,j_2+1}\\
    	 & + (\cdots)\sqrt{n_2}\sqrt{j_1+1}\delta_{m_1,n_1}\delta_{m_2,n_2-1}\delta_{k_1,j_1+1}\delta_{k_2,j_2}\\
         & + (\cdots)\sqrt{n_2}\sqrt{j_2+1}\delta_{m_1,n_1}\delta_{m_2,n_2-1}\delta_{k_1,j_1}\delta_{k_2,j_2+1}.
    \end{split}
\end{gather}

Two transition processes can be distinguished: Process 1, represented by the first and the last terms, are transitions between the 'steps' of the same 'ladder' of one of the harmonic oscillators. Process 2, represented by the second and third terms, are transitions from one Harmonic-ladder to the other. If one examines the complex exponent in the master equation  (Eq.(\ref{EqAPPMasterEqShiftTime})), $\exp[i(\Delta_{mn}-\Delta_{jk})t]$, one sees process Process 1 gives exponents of unity. However, Process 2 gives exponents of the form $\exp[\pm i(\varepsilon_1-\varepsilon_2)t]$, which are a quickly oscillating terms. It is noted, that neglecting process 2, in general does influence the proper description of the dynamics of the system\cite{settineri2018dissipation}. However, in view of the semi-classical approximation, the Rotating wave approximation is applied and process 2 transitions are neglected in the calculations of this study.

\section{Dissipator in the unrotated and unshifted frame.}\label{APPDissipatorConstants}
This section outlines how the rotated dissipator is transformed into the unrotated frame. This section presents the final result of a diagonalized dissipator and defines the associated constants. 

The diagonalized destruction operators ($\hat{d}$), expressed shifted photonic operators ($\hat{c}$) and the rotated Holstein-Primakoff operators ($\hat{b}$), are of the form of Eq.(\ref{EqAPPOpsddIncb}) or $\hat{d} = \alpha \hat{c} + \beta\hat{c}^\dagger + \gamma \hat{b} + \delta \hat{b}^\dagger$. After performing the procedure outlined in section \ref{SectionDissipatorOfpqS}, the dissipator in the rotated frame expressed in the rotated operators becomes: 
\begin{gather}\label{EqDissipatorRotatedDressed}
    \begin{split}
	       \mathcal{D}[\hat{\tilde{q}}] = & - \frac{\alpha^2 - \beta^2}{2}\hat{\tilde{q}} + \sqrt{\frac{1}{\omega S}}\frac{(\beta-\alpha)(\delta + \gamma)}{2}\hat{J}_x\\
	       \mathcal{D}[\hat{\tilde{p}}] = & - \frac{\alpha^2 - \beta^2}{2}\hat{\tilde{p}} + \frac{(\alpha+\beta)(\delta-\gamma)}{2}\sqrt{\frac{\omega}{S}}\hat{J}_y \\
	       \mathcal{D}[\hat{J}_x] = &- \frac{\gamma^2-\delta^2}{2}\hat{J}_x + \frac{(\delta - \gamma)(\alpha + \beta)}{2}\sqrt{\frac{\omega S}{1}} \hat{\tilde{q}}\\
	       \mathcal{D}[\hat{J}_y] = &  - \frac{\gamma^2-\delta^2}{2}\hat{J}_y + \frac{(\delta + \gamma)(\beta - \alpha)}{2}\sqrt{\frac{S}{\omega}} \hat{\tilde{p}}\\
	       \mathcal{D}[\hat{J}_z] = & (\gamma^2-\delta^2)(S-\hat{J}_z) + \frac{(\alpha+\beta)(\gamma-\delta)}{2}\sqrt{\frac{\omega}{S}}\hat{\tilde{q}}\hat{J}_x\\
           &     + \frac{(\alpha-\beta)(\gamma+\delta)}{2}\sqrt{\frac{1}{\omega S}} \hat{\tilde{p}} \hat{J}_y.
    \end{split}
\end{gather}
Where the constants $\alpha, \beta, \gamma, \delta$ are the coefficients of the diagonalized annihilation operator ($\hat{d} = \alpha \hat{c} + \beta \hat{c}^\dagger + \gamma \hat{b} + \delta \hat{b}^\dagger$).

To transform from the rotated and shifted frame to the normal frame, one can use the transformations of Eq.(\ref{EqAPPRotTrans}) combined with the linearity of the dissipator to find: 
\begin{gather}\label{EqRotationDissipator}
    \begin{split}
    	\mathcal{D}[\hat{q}] = & \mathcal{D}[\hat{\tilde{q}}]\\
    	\mathcal{D}[\hat{p}] = & \mathcal{D}[\hat{\tilde{p}} + p_0]\\
    	\mathcal{D}[\hat{S}_x] = & \mathcal{D}[\hat{J}_x]\\
    	\mathcal{D}[\hat{S}_y] = & \cos(\theta)\mathcal{D}[\hat{J}_y] + \sin(\theta)\mathcal{D}[\hat{J}_z]\\
    	\mathcal{D}[\hat{S}_z] = & -\sin(\theta)\mathcal{D}[\hat{J}_y] + \cos(\theta)\mathcal{D}[\hat{J}_z].
    \end{split}
\end{gather}

Plugging in the equations from Eq.(\ref{EqDissipatorRotatedDressed}) into Eq.(\ref{EqRotationDissipator}) to find the dissipators in the normal ('unshifted and unrotated') frame, one finds:
\begin{gather}\label{EqDissipatorNormalDressed}
    \begin{split}
            \mathcal{D}_{d_{1,2}}[\hat{q}] = & - A_{1,2} \hat{q} + B_{1,2} \hat{S}_x\\
        	\mathcal{D}_{d_{1,2}}[\hat{p}] = & - A_{1,2} \left(\hat{p}-p_0\right) + C_{1,2} \cos(\theta)\hat{S}_y - C_{1,2}\sin(\theta)\hat{S}_z\\
        	\mathcal{D}_{d_{1,2}}[\hat{S}_x] = & - D_{1,2} \hat{S}_x + E_{1,2} \hat{q}\\
        	\mathcal{D}_{d_{1,2}}[\hat{S}_y] = & F_{1,2}\cos(\theta)(\hat{p}-p_0) + C_{1,2}\sin(\theta)\hat{q}\hat{S}_x + 2D_{1,2}\sin(\theta)S\\
        		& - \left[D_{1,2}(1+\sin^2(\theta)) - \frac{B_{1,2}}{2}\sin(2\theta)(\hat{p}-p_0)\right]\hat{S}_y\\
        		& + \left[-B_{1,2}\sin^2(\theta)(\hat{p}-p_0) - \frac{D_{1,2}}{2}\sin(2\theta)\right]\hat{S}_z\\
        	\mathcal{D}_{d_{1,2}}[\hat{S}_z] = & 2D_{1,2}\cos(\theta)S - F_{1,2}\sin(\theta)(\hat{p}-p_0) +  C_{1,2}\cos(\theta)\hat{q}\hat{S}_x\\
        		& - \left[\frac{D_{1,2}}{2}\sin(2\theta) - B_{1,2}\cos^2(\theta)(\hat{p}-p_0))\right]\hat{S}_y\\
        		& + \left[-\frac{B_{1,2}}{2}\sin(2\theta)(\hat{p}-p_0) - D_{1,2}\left(1+\cos^2(\theta)\right)\right]\hat{S}_z,
    \end{split}
\end{gather}

where $-p_0=\pm gS_{y}^{SR}$, $\sin(\theta) = S_y^{SR}/S$, $\cos(\theta) = S_z^{SR}/S$. The other constants are defined as:
\begin{gather}
    \begin{split}
        & 2A_1 = 2D_2 = \cos^2(\chi)\\
        & 2A_2 = 2D_1 = \sin^2(\chi)\\
        & 2B_{1,2} = \mp \sin(\chi)\cos(\chi)\frac{1}{\omega}\sqrt{\frac{F}{S}}\\
        & 2C_{1,2} = \mp \sin(\chi)\cos(\chi) \frac{\omega}{\sqrt{SF}}\\
        & 2E_{1,2} = \mp \sin(\chi)\cos(\chi) \omega \sqrt{\frac{S}{F}}\\
        & 2F_{1,2} = \mp \sin(\chi)\cos(\chi) \frac{\sqrt{SF}}{\omega}
    \end{split}
\end{gather}

\bibliographystyle{apsrevlong_no_issn_url.tex}

\bibliography{bibliography}

\end{document}